\documentstyle[aps,twocolumn,epsf]{revtex}
\begin{document}
\title{\hspace{7cm} {\em to appear in Phys. Rev. B}\\
\vskip 0.25cm
Phase dependent differential thermopower of $SND$ junctions:
Pair-breaking effects and Gaussian fluctuations}
\author{Sergei Sergeenkov\cite{byline} and Marcel Ausloos}
\address{SUPRAS, Institute of Physics, University of Liege,
B-4000, Liege, Belgium}
\date{\today}
\draft
\maketitle
\begin{abstract}
We start with revisiting our previous results on thermoelectric response of
$SNS$ configuration in a $C$-shaped
$Bi_xPb_{1-x}Sr_2CaCu_2O_y$ sample in order to include strong fluctuation
effects. Then, by appropriate
generalization of the Ginzburg-Landau theory based on admixture
of $s$-wave (S) and $d$-wave (D) superconductors, we consider a differential
thermoelectric power (TEP) of $SND$ junction. In addition to its strong
dependence on the relative phase $\theta =\phi _s-\phi _d$ between the two
superconductors, two major effects are shown to influence the behavior of the
predicted TEP. One, based on the chemical imbalance at $SD$ interface,
results in a pronounced
maximum of the TEP peak near $\theta =\pi /2$ (where the so-called $s+id$
mixed pairing state is formed) for two identical superconductors with
$T_{cd}=T_{cs}\equiv T_c$. Another effect, which should manifest itself at
$SD$ interface comprising an $s$-wave low-$T_c$ superconductor and a $d$-wave
high-$T_c$ superconductor with $T_{cd}\neq T_{cs}$, predicts
$S_p\propto T_{cd}-T_{cs}$ for the TEP peak value. The experimental
conditions under which the predicted behavior of the induced
differential TEP can be measured are discussed.
\end{abstract}

\narrowtext

\section{Introduction}
During the last few years the order parameter symmetry has been
one of the intensively debated issues in the field of high-$T_c$
superconductivity (HTS). A number of experiments points to its
$d_{x^2-y^2}$-wave character~\cite{1}. Such an unconventional
symmetry of the order parameter has also important implications for the
Josephson physics
because for a $d$-wave (D) superconductor the Josephson coupling is subject
to an additional phase dependence caused by the internal phase
structure of the wave function.  The phase properties of
the Josephson effect have been discussed within the framework of the
generalized Ginzburg-Landau (GL)~\cite{2} as well as the tunneling
Hamiltonian approach~\cite{3}. It was found~\cite{4} that the
current-phase
relationship depends on the mutual orientation of the two coupled
superconductors and their interface.  This property is the basis of
all the phase sensitive experiments probing the order parameter symmetry.
In particular, it is possible to create multiply connected $d$-wave
superconductors which generate half-integer flux quanta as observed in
experiments~\cite{5}.
Various interesting phenomena occur in interfaces of $d$-wave
superconductors. For example, for an interface to a normal metal a bound
state appears at zero energy giving rise to a zero-bias anomaly in the
$I$-$V$-characteristics of quasiparticle tunneling~\cite{6,7}
while in such an interface to an $s$-wave (S) superconductor the energy
minimum
corresponds to a Josephson phase different from $0$ or $\pi$.
By symmetry, a small $s$-wave component always coexists with a predominantly
$d$-wave order parameter in an orthorhombic superconductor such as $YBCO$,
and changes its sign across a twin boundary~\cite{8,9}.
Besides, the $s$-wave and $d$-wave order parameters can form a complex
combination, the so-called $s\pm id$-state which is
characterized by a local breakdown of time reversal symmetry
${\cal T}$ either near surfaces~\cite{10,11,12,13} or near the twin
boundaries represented by tetragonal regions with a reduced
chemical potential~\cite{14}. Both scenarios lead to
a phase difference of $\pm \pi/2$, which corresponds to two
degenerate states~\cite{15,16}. Moreover,
the relative phase oscillations between two condensates with
different order parameter symmetries could manifest themselves through
the specific collective excitations ("phasons")~\cite{17}.

At the same time, a rather sensitive differential technique to probe
sample inhomogeneity for temperatures just below $T_c$, where phase
slippage events play an important role in transport characteristics
has been proposed~\cite{18} and successfully applied~\cite{19} for
detecting small changes in thermoelectric power (TEP) of a specimen due to
the
deliberate insertion of a macroscopic $SNS$ junction made of a normal-metal
layer $N$, used to force pair breaking of the superconducting
component when it flows down the temperature gradient.
Analysis of the thermoelectric
effects provides reasonable estimates for such important physical parameters
as the Fermi energy, Debye temperature, interlayer spacing etc.
In particular, a carrier-type-dependent thermoelectric response of such
a $SNS$ configuration in a $C$-shaped $Bi_xPb_{1-x}Sr_2CaCu_2O_y$ sample
has been registered and its temperature behavior
below $T_c$ has been explained within the framework of GL
theory~\cite{19}.

In the present paper, we consider theoretically the behavior of induced TEP
at $NS$, $ND$, and $SD$ interfaces and discuss its possible implications for
the above-mentioned type of
experiments. The paper is organized as follows. In Section II
we briefly review the experimental results for $SNS$ configuration (with
both holelike and electronlike carriers of the normal-metal $N$ insert)
and present a theoretical interpretation of these results,
based on GL free energy functional, both below and above $T_c$. The crucial
role of the pair-breaking  effects (described via the chemical balance
$\Delta \mu $ between the quasiparticles and Cooper pairs) in understanding
the observed phenomena is emphasized.
In Section III, extending the early suggested~\cite{11,14} GL theory of
an admixture of $s$-wave and $d$-wave superconductors to incorporate
strong pair-breaking  effects, we calculate the differential thermopower
$\Delta S$ of $SND$ configuration near $T_c$. The main
theoretical result of this Section is the prediction of a rather specific
dependence of $\Delta S$ on relative phase shift $\theta  =\phi _s-\phi _d$
between the two superconductors.
Two independent mechanisms contributing to the peak value
$S_p(\theta )=\Delta S(T_c,\theta )$ of the differential thermopower
are discussed. One, based on
the chemical balance between $S$ and $D$ superconductors at an $SD$
interface (and responsible for charge-related interference effect), is
discussed in Section IIIA. It results in a
pronounced maximum of the peak $S_p(\theta )$ near
$\theta =\pi /2$ (the so-called $s+id$ mixed pairing state) for two identical
superconductors with $T_{cd}=T_{cs}\equiv T_c$. This mechanism can be
realized, e.g., in a $d$-wave orthorhombic sample
(like $YBCO$) with twin boundaries which are represented by tetragonal
regions of variable width, with a reduced chemical potential.
Another mechanism (discussed in Section IIIB), which is active in the absence
of the normal-metal layer, takes place when
two different superconductors with $T_{cd}\neq T_{cs}$ are used to form
an $SD$ interface. This situation can be realized for an $s$-wave low-$T_c$
superconductor (like $Pb$) and a $d$-wave high-$T_c$ superconductor
(like orthorhombic $YBCO$) and is shown to yield $S_p(\theta )
\propto T_{cd}-T_{cs}$ for predicted TEP peak value.

\section{$SNS$ configuration revisited}

\subsection{Experimental setup and main results}

Before turning to the main subject of the present paper, let us
briefly review the previous results concerning a
thermoelectric response of $SNS$ configuration in a $C$-shaped
$Bi_xPb_{1-x}Sr_2CaCu_2O_y$ sample (see Ref.19 for details).
The sample geometry used is sketched in Fig.1, where the contact
arrangement and
the position of the sample with respect to the temperature gradient
$\nabla _xT$ is shown as well.
Two cuts are inserted at $90^o$ to each other
into a ring-shaped superconducting sample. The first cut lies parallel
to the applied temperature gradient serving to define a vertical symmetry
axis. The second cut lies in the middle of the right wing, normal to the
symmetry axis, separating two similar superconductors with $S'=S''=S$ or $D$
and completely interrupting the passage of supercurrents in
this wing. The passage of any normal component of current
density is
made possible by filling up the cut with a normal metal $N$.
The carrier type of the normal-metal insert
$N$ was chosen to be either an electronlike $N_e$ (silver) or holelike $N_h$
(indium).
Thermal voltages resulting from the same temperature gradient acting on
both continuous and normal-metal-filled halves of the sample were detected
as a function of temperature around $T_c$.
\begin{figure}[htb]
\epsfxsize=8cm
\centerline{\epsffile{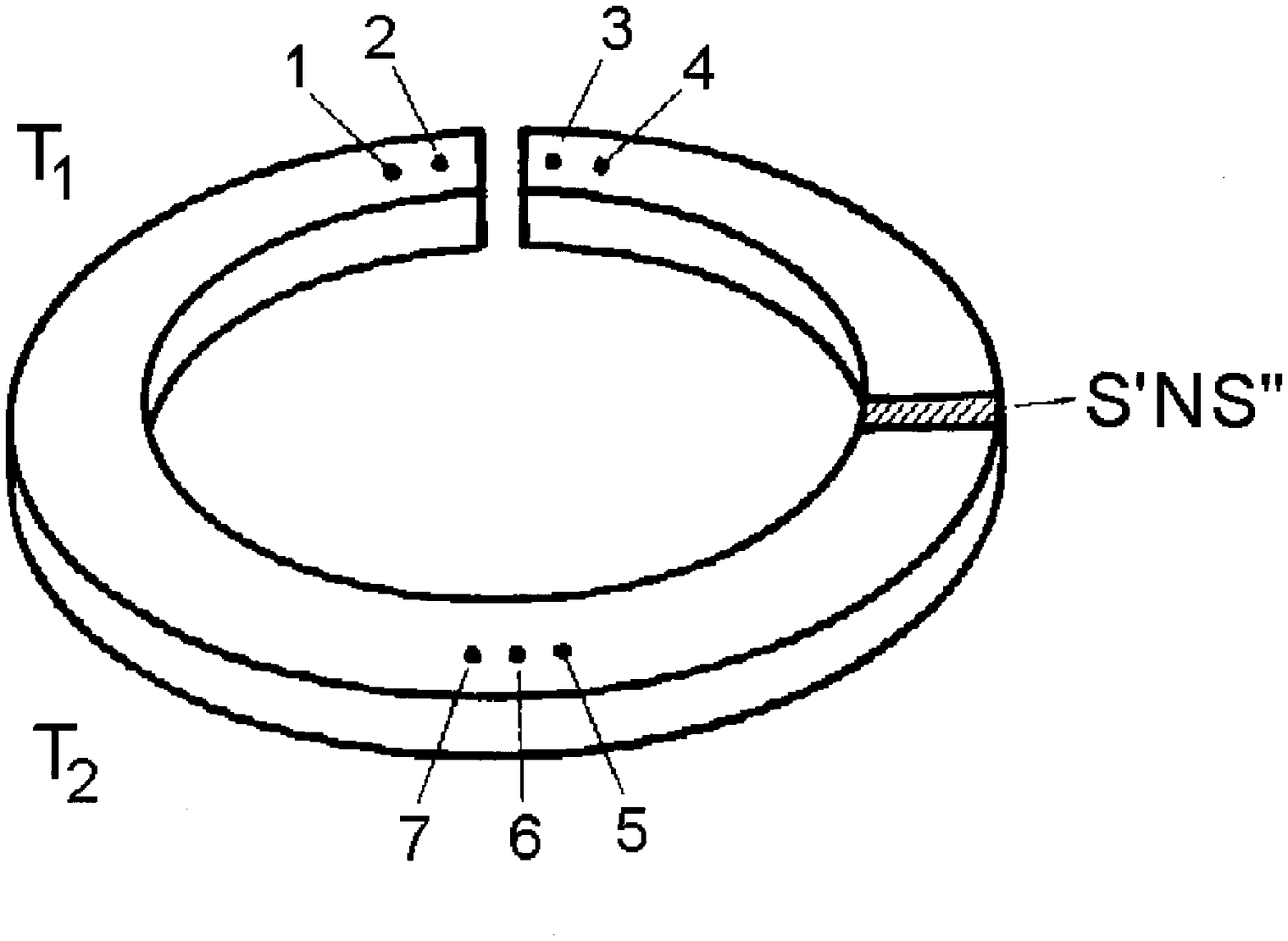} }
\caption{Schematic view of the sample geometry with $S'NS''$-junction and
contacts configuration. Here $S'$ and $S''$ stand for $s$-wave and/or
$d$-wave type superconductor. The thermopowers $S_R$ and $S_L$ result from the
thermal voltages detected by the contact pairs $4-5$ and $1-7$, respectively.}
\end{figure}
The measured difference between
the thermopowers of the two halves $\Delta S=S_R-S_L$ was found to
approximately follow the linear dependence
\begin{equation}
\Delta S(T)\simeq S_p\pm B^{\pm}(T_c-T),
\end{equation}
with slopes $B^{-}$ and $B^{+}$ defined for $T<T_c$ and $T>T_c$,
respectively. Here $S_p=\Delta S(T_c)$ is the peak
value of $\Delta S(T)$
at $T=T_c$. The best fit of the experimental data with the
above equation yields the following values for silver (Ag) and indium (In)
inserts, respectively (see Fig.2): (i) $S_p(Ag)=-0.26\pm 0.01 \mu V/K$,
$B^{-}(Ag)=-0.16\pm 0.1 \mu V/K^2$,
$B^{-}(Ag)/B^{+}(Ag)=1.9\pm 0.1$;
(ii) $S_p(In)=0.83\pm 0.01 \mu V/K$, $B^{-}(In)=
0.17\pm 0.1 \mu V/K^2$, $B^{-}(In)/B^{+}(In)=2.1\pm 0.1$.

\subsection{Interpretation}

It is important to mention that, unlike the case of mixed $SND$ configuration
(considered in Section III), the suggested interpretation of the current
experimental results for $SNS$ configuration does not involve the phase of
the order parameter and hence is not sensitive (at least near $T_c$) to the
pairing symmetry of the two superconductors $S'$ and $S''$.
To describe the observed behavior of the differential
TEP both below and above $T_c$, we can roughly present it in a two-term
contribution form~\cite{19,20}
\begin{equation}
\Delta S(T)=\Delta S_{av}(T)+\Delta S_{fl}(T),
\end{equation}
where the average term $\Delta S_{av}(T)$ is assumed to be non-zero only
below $T_c$ (since in the normal state the TEP of HTS is found to be
very small~\cite{21,22}) while the fluctuation term $\Delta S_{fl}(T)$
should contribute to the observable $\Delta S(T)$ for $T\simeq T_c$.
In what follows, we shall discuss these two contributions separately within
a mean-field theory approximation.
\begin{figure}[htb]
\epsfxsize=8cm
\centerline{\epsffile{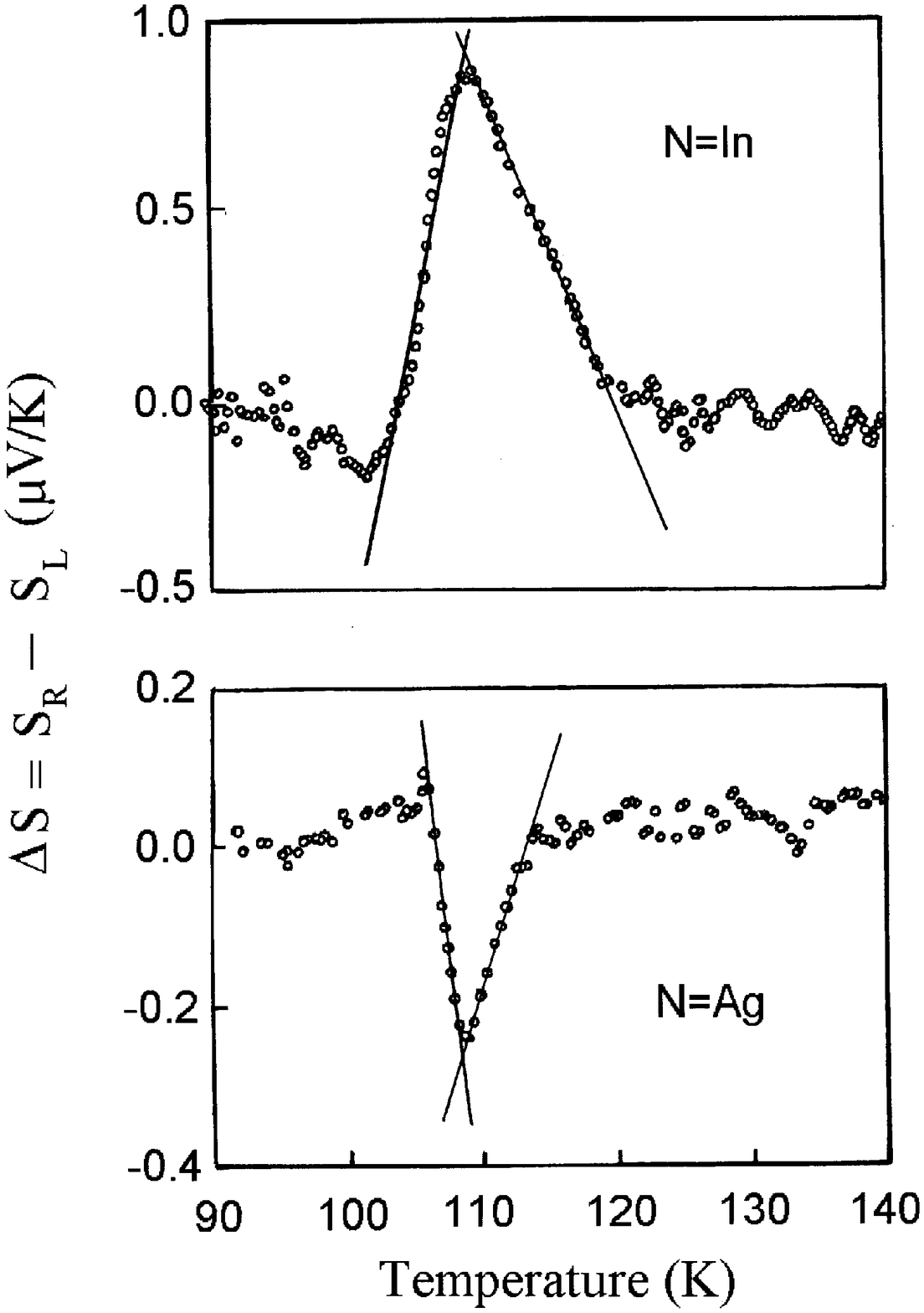} }
\caption{The temperature dependence of the observed differential thermopower
of $SNS$ configuration defined by Eq.(1). The upper (lower) part of the
picture refers to In (Ag) normal-metal insert in the right wing (see Fig.1).
The asymmetric curved triangle shapes are approximated by linear shapes
produced by the linear fit to the data points (see the text for details).}
\end{figure}

\subsubsection{Mean value of the differential thermopower:
$\Delta S_{av}(T)$}

Assuming that the net result of the normal-metal insert is to break up
Cooper pairs that flow toward the hotter end of the sample and to produce
holelike (In) or electronlike (Ag) quasiparticles, we can write the
difference in the generalized GL free energy functional $\Delta {\cal G}$ of
the right and left halves of the $C$-shaped sample as
\begin{equation}
\Delta {\cal G} [\psi ]=\Delta {\cal F}[\psi ]+\Delta \mu |\psi |^2,
\end{equation}
where
\begin{equation}
\Delta {\cal F}[\psi ]\equiv {\cal F}_R-{\cal F}_L=a(T)|\psi |^2+
\frac{\beta}{2}|\psi |^4
\end{equation}
and
\begin{equation}
\Delta \mu \equiv \mu _R-\mu _L.
\end{equation}
Here $\psi =|\psi |e^{i\phi}$ is the superconducting order parameter,
$\Delta \mu$ accounts for the chemical balance between quasiparticles and
Cooper pairs; $a(T)=\alpha (T-T_c)$ and the GL parameters
$\alpha$ and $\beta$ are related to the critical temperature $T_c$,
zero-temperature BCS gap $\Delta _0=1.76k_BT_c$, the in-plane Fermi energy
$E_F^{ab}=p_F^2/m_{ab}^{*}$, and the total particle number density $n$ as
$\alpha =\beta n/2T_c=(4\Delta _0k_B/E_F^{ab})(m_e/m_{ab}^{*})$. In fact, in
layered superconductors, $\Delta \mu \simeq
E_F^c\simeq J_c^2/E_F^{ab}$, where $E_F^c=E_F^{ab}/\gamma _m^2$ is the
out-of-plane Fermi energy and $J_c$ the interlayer coupling energy within
the Lawrence-Doniach model ($\gamma _m=m_c^{*}/m_{ab}^{*}$ is the
mass anisotropy ratio, and $m_{ab}^{*}\simeq 8m_e$ for this material).

As usual, the equilibrium state of such a system is determined from the
minimum energy condition $\partial {\cal G}/\partial |\psi |=0$ which
yields for $T<T_c$
\begin{equation}
|\psi _0|^2=\frac{\alpha (T_c-T)-\Delta \mu}{\beta}
\end{equation}
Substituting $|\psi _0|^2$ into Eq.(3) we obtain for the average free
energy density
\begin{equation}
\Delta \Omega (T)\equiv \Delta {\cal G} [\psi _0]=-
\frac{[\alpha (T_c-T)-\Delta \mu ]^2}{2\beta}
\end{equation}
In turn, the difference of thermopowers $\Delta S(T)$ can be
related to the
corresponding difference of transport entropies $\Delta \sigma \equiv
-\partial \Delta \Omega /\partial T$ as $\Delta S(T)=\Delta \sigma (T)/nq$,
where $q$ is the charge of the quasiparticles.
Thus finally the mean value of thermopower associated with a pair-breaking
event reads (below $T_c$)
\begin{equation}
\Delta S_{av}(T)=S_{p,av}-B_{av}(T_c-T),
\end{equation}
with
\begin{equation}
S_{p,av}=\frac{\Delta \mu}{2qT_c},
\end{equation}
and
\begin{equation}
B_{av}=\frac{\Delta _0k_B}{2qE_F^{ab}T_c}.
\end{equation}
Before we proceed to compare the above theoretical findings with the
available experimental
data (see Fig.2), we first have to estimate the corresponding fluctuation
contributions to the observable TEP difference, both above and below $T_c$.

\subsubsection{Mean-field Gaussian fluctuations of the differential
thermopower: $\Delta S_{fl}(T)$}

The influence of superconducting fluctuations on transport properties of HTS
(including TEP and electrical conductivity) has been extensively
studied for the past few years
(see, e.g.,~\cite{23,24,25,26,27,28,29,30,31,32} and further references
therein). In particular, it was found that the
fluctuation-induced behavior may extend to temperatures more than
$10K$ higher than the respective $T_c$.
Let us consider now the region near $T_c$ and discuss the Gaussian
fluctuations of the pair-breaking-induced differential TEP
$\Delta S_{fl}(T)$.
Recall that according to the theory of Gaussian fluctuations~\cite{33},
the fluctuations of any observable, which is conjugated to the order
parameter $\psi$ (such as heat
capacity, susceptibility, etc) can be presented in terms of the statistical
average of the square of the fluctuation amplitude $<(\delta \psi )^2>$ with
$\delta \psi =\psi -\psi _0$. Then the differential TEP above $(+)$ and
below $(-)$ $T_c$ have the form of
\begin{equation}
\Delta S_{fl}^{\pm}(T)=A<(\delta \psi )^2>_{\pm},
\end{equation}
where
\begin{equation}
<(\delta \psi )^2>=\frac{1}{Z}\int d|\psi |(\delta \psi )^2 e^{-\Sigma [\psi ]}.
\end{equation}
Here
$Z=\int d|\psi |e^{-\Sigma [\psi ]}$ is the partition function with
$\Sigma [\psi ]\equiv (\Delta {\cal G} [\psi ]-
\Delta {\cal G} [\psi _0])/k_BT$. $A$ is a coefficient to be defined below.
Expanding the free energy density functional $\Delta {\cal G} [\psi ]$
\begin{equation}
\Delta {\cal G} [\psi ]\approx \Delta {\cal G} [\psi _0]+
\frac{1}{2}\left[ \frac{\partial ^2\Delta {\cal G}}
{\partial \psi ^2}\right ]_{|\psi |=|\psi _0|}\!(\delta \psi )^2,
\end{equation}
around the mean value of the order parameter $\psi _0$, which is defined as a
stable solution of equation $\partial {\cal G}/\partial |\psi |=0$ we can
explicitly calculate the Gaussian integrals.
Due to the fact that $|\psi _0|^2$ is given by Eq.(6) below $T_c$ and
vanishes at $T\ge T_c$, we obtain finally
\begin{equation}
\Delta S_{fl}^{-}(T)=\frac{Ak_BT_c}{4\alpha (T_c-T)-
4\Delta \mu }, \qquad T\le T_c
\end{equation}
and
\begin{equation}
\Delta S_{fl}^{+}(T)=\frac{Ak_BT_c}{2\alpha (T-T_c)+
2\Delta \mu }, \qquad T\ge T_c
\end{equation}
As we shall see below, for the experimental range of parameters under
discussion, $\Delta \mu (E_F/\Delta _0)\gg k_B|T_c-T|$. Hence, with a good
accuracy we can approximate Eqs.(14) and (15) as follows
\begin{equation}
\Delta S_{fl}^{\pm}(T)\simeq S_{p,fl}^{\pm}\pm B_{fl}^{\pm}(T_c-T),
\end{equation}
where
\begin{equation}
S_{p,fl}^{-}=-\frac{Ak_BT_c}{4\Delta \mu},
\qquad B_{fl}^{-}=\frac{Ak_BT_c\alpha}{(2\Delta \mu )^2},
\end{equation}
and
\begin{equation}
S_{p,fl}^{+}=-2S_{p,fl}^{-}, \qquad B_{fl}^{+}=2B_{fl}^{-}.
\end{equation}
Furthermore, it is quite reasonable to assume that $S_p^{-}=S_p^{+}\equiv
S_p$, where $S_{p}^{-}=S_{p,av}+S_{p,fl}^{-}$ and $S_{p}^{+}=S_{p,fl}^{+}$.
Then the above equations
bring about the following explicit expression for the constant
parameter $A$, namely $A=(4\Delta \mu /3k_BT_c)S_{p,av}$. This in turn leads
to the following expressions for the fluctuation and total contributions to
peaks and slopes through their average counterparts (see Eqs.(9) and (10)):
$S_{p,fl}^{+}=S_p=(2/3)S_{p,av}$, $S_{p,fl}^{-}=-(1/3)S_{p,av}$,
$B_{fl}^{-}=(1/3)B_{av}$, $B_{fl}^{+}=
(2/3)B_{av}$, $B^{-}=B_{av}+B_{fl}^{-}=(4/3)B_{av}$, and $B^{+}=B_{fl}^{+}=
(2/3)B_{av}$. Thus, in agreement with the observations, $B^{-}/B^{+}=2$
independent of the carrier type of the normal-metal insert.
Let us proceed to discuss separately the case of $In$ and $Ag$ inserts.

\paragraph{$N=In$ (holelike metal insert).}

In this case, the principal carriers are holes, therefore $q=+e$.
Let the holelike quasiparticle chemical potential (measured relative to
the Fermi level of the free-hole gas) be positive, then $\mu _q=+\mu$
and $\Delta \mu =\mu +2\mu =3\mu$ (two holes come from condensate and one
hole is brought by normal metal). Therefore,
for this case Eq.(1) takes the form
\begin{equation}
\Delta S^h(T)=S_p(In)\pm B^{\pm}(In)(T_c-T),
\end{equation}
where
\begin{equation}
S_p(In)=\left (\frac{k_B}{e}\right )\left (\frac{\mu}{k_BT_c}\right ),
\end{equation}
and
\begin{equation}
B^{-}(In)=\frac{2\Delta _0k_B}{3eE_F^{ab}T_c},
\qquad B^{+}(In)=\frac{1}{2}B^{-}(In).
\end{equation}

\paragraph{$N=Ag$ (electronlike metal insert).}

The principal carriers in this case are electrons, therefore $q=-e$. The
electronlike quasiparticle chemical potential $\mu _q=-\mu$.
Then $\Delta \mu =-\mu +2\mu =\mu$ (plus one electron means minus one hole).
For this case Eq.(1) takes the form
\begin{equation}
\Delta S^e(T)=S_p(Ag)\pm B^{\pm}(Ag)(T_c-T),
\end{equation}
where
\begin{equation}
S_p(Ag)=-\left (\frac{k_B}{e}\right )\left (\frac{\mu}{3k_BT_c}\right ),
\end{equation}
and
\begin{equation}
B^{-}(Ag)=-\frac{2\Delta _0k_B}{3eE_F^{ab}T_c},
\qquad B^{+}(Ag)=\frac{1}{2}B^{-}(Ag).
\end{equation}
By comparing the obtained theoretical expressions with the above-mentioned
experimental findings for the slopes $B^{\pm}$ and the
peak $S_p$ values for the two normal-metal inserts (see Fig.2), we can
estimate the order of magnitude of the in-plane Fermi energy $E_F^{ab}$ and
interlayer
coupling energy $J_c$. The result is: $E_F^{ab}\simeq 0.16eV$ and
$J_c\simeq 4meV$, in reasonable agreement
with the other known estimates of these parameters~\cite{30}.
In turn, using these parameters (along with the critical temperature), we find
that $J_c^2/k_B\Delta _0\simeq 100K$. This justifies the use of the
linearized Eq.(16) for the temperature interval $|T_c-T|\ll 100K$.
As is seen in Fig.2, the observed differential TEP practically disappears
already for $|T_c-T|\ge 10K$.
Moreover, as it follows from Eqs.(20) and (23), the calculated ratio for peaks
$|S_p(In)/S_p(Ag)|=3$ is very close to the corresponding experimental value
$|S_p^{exp}(In)/S_p^{exp}(Ag)|=3.2\pm 0.2$ observed by Gridin et
al~\cite{19}. Finally, as it follows from the above analysis, the calculated
slopes $B^{-}$ below $T_c$ for the two metal inserts coincide with each other,
namely $B^{-}(In)=-B^{-}(Ag)$, and are twice their counterparts above $T_c$,
i.e., $B^{-}(In)=2B^{+}(In)$ and $B^{-}(Ag)=2B^{+}(Ag)$, in a good agreement
with the observations.
It is worthwhile to note that a very similar behavior of the induced TEP
(including peaks and slopes both above and below $T_c$)
has been observed in strong applied magnetic fields~\cite{20}. In fact,
replacing the chemical potentials difference $\Delta \mu$
(responsible for pair-breaking  effects in $SNS$ junction) in the above
equations by $\mu _BH$ term (where $\mu _B$ is the Bohr magneton and $H$ the
applied magnetic field) we recover most
of the formulas presented in Ref.20 where magneto-TEP of
$Bi_2Sr_2CaCu_2O_y$ superconductors was studied.

\section{$SND$ configuration: prediction}

Since Eqs.(3)-(5) do not depend on the phase of the order parameter, they
will preserve their form for a $DND$ junction (created by two $d$-wave
superconductors, $S'=S''=D$, see Fig.1) as well, bringing about
the results similar to that given by Eqs.(8)-(10). It means that the
experimental method
under discussion (and its interpretation) can not be used to tell
$SNS$ and $DND$ configurations apart, at least for
temperatures close to $T_c$. As for low enough temperatures, the situation
may change drastically due to a markedly different behavior of $s$-wave and
$d$-wave order parameters at $T\ll T_c$ (where the node structure begins to
play an important role). As we shall show below, this method, however,
is quite sensitive to the mixed $SND$ configuration (when $S'=S$
has an $s$-wave symmetry while $S''=D$ is of a $d$-wave symmetry type, see
Fig.1) predicting a
rather specific relative phase ($\theta =\phi _s-\phi _d$) dependencies of
both the slope $B(\theta )$ and peak $S_p(\theta )$ of the observable
thermopower difference $\Delta S(T,\theta )$.

Following Feder et al~\cite{14}, who incorporated chemical potential effects
near twin boundaries into the approach suggested by Sigrist et al~\cite{11},
we can represent
the generalized GL free energy functional $\Delta {\cal G}$ for $SND$
configuration of the $C$-shaped sample in the following form
\begin{equation}
\Delta {\cal G} [\psi _s,\psi _d]=\Delta {\cal G} [\psi _s]+
\Delta {\cal G} [\psi _d]+\Delta {\cal G}_{int},
\end{equation}
where
\begin{equation}
\Delta {\cal G} [\psi _{s}]=A_s(T)|\psi _{s}|^2+
\frac{\beta _s}{2}|\psi _{s}|^4,
\end{equation}
\begin{equation}
\Delta {\cal G} [\psi _{d}]=
A_d(T)|\psi _{d}|^2+\frac{\beta _d}{2}|\psi _{d}|^4,
\end{equation}
and
\begin{eqnarray}
\Delta {\cal G}_{int}=\gamma _1|\psi _s|^2|\psi _d|^2+
\frac{\gamma _2}{2}(\psi _s^{*2}\psi _d^2+\psi _s^2\psi _d^{*2})&&\\ \nonumber
-2\delta _1|\psi _s||\psi _d|-
\delta _2(\psi _s^{*}\psi _d+\psi _s\psi _d^{*}).&&
\end{eqnarray}
Here $\psi _n=|\psi _n|e^{i\phi _n}$ is the $n$-wave order parameter
($n=\{s,d\}$); $A_n(T)=a_n(T)+\Delta \mu _n$ where
$a_n(T)=\alpha _n(T-T_{cn})$ with the corresponding parameters $\alpha _n$,
$\beta _n$, $T_{cn}$, and $\Delta \mu _n$ for $s$-wave and $d$-wave
symmetries.

An equilibrium state of such a mixed system is determined from the
minimum energy conditions $\partial {\cal G}/\partial |\psi _s|=0$
and $\partial {\cal G}/\partial |\psi _d|=0$ which result in the following
system of equations for the two equilibrium order parameters
$\psi _{s0}$ and $\psi _{d0}$
\begin{eqnarray}
&&A_s|\psi _{s0}|+\beta _s|\psi _{s0}|^3+
\Gamma (\theta )|\psi _{s0}||\psi _{d0}|^2=\Delta (\theta )|\psi _{d0}|\\
&&A_d|\psi _{d0}|+\beta _d|\psi _{d0}|^3+
\Gamma (\theta )|\psi _{d0}||\psi _{s0}|^2=\Delta (\theta )|\psi _{s0}|
\end{eqnarray}
where we introduced relative phase $\theta =\phi _s-\phi _d$ dependent
parameters
\begin{eqnarray}
&&\Gamma (\theta )=\gamma _1+\gamma _2\cos 2\theta \\ \nonumber
&&\Delta (\theta )=\delta _1+\delta _2\cos \theta
\end{eqnarray}
Notice that unlike chemical potentials difference $\Delta \mu _n$ (which is
responsible for pair-breaking  effects in $SND$ junction due to the
normal-metal insert), the interference terms $\delta _{1,2}$ describe the
chemical balance between $s$-wave and $d$-wave superconductors
at $SD$ interface in the absence of a normal-metal layer. Therefore,
the effects due to $\Delta \mu _n\neq 0$ should be distinguished from the
interference effects due to $\Delta (\theta )\neq 0$. The latters are
generically close to the interference effects between the two condensates
and are described by the $\Gamma (\theta )$ term.
Notice also that the $\Delta (\theta )$ term favors $\theta =l\pi $ ($l$
integer),
while the $\Gamma (\theta )$ term favors $\theta =n\pi /2$ ($n=1,3,5\ldots$)
which corresponds to a ${\cal T}$-violating phase~\cite{14}.
In principle, we can resolve the above system (given by Eqs.(29)-(31)) and
find $\psi _{n0}$ for arbitrary set of parameters $\alpha _n$,
$\beta _n$, and $T_{cn}$. For simplicity, in what follows
we restrict our consideration to the two limiting cases which are of
the most importance for potential applications.

\subsection{Twin boundaries in orthorhombic $d$-wave superconductors}

\subsubsection{Mean value of the differential thermopower:
$\Delta S_{av}(T,\theta )$}

First, let us consider the case of similar superconductors comprising the
$SND$ junction with $|\psi _{s0}|=
|\psi _{d0}|\equiv |\psi _0|$, $\alpha _s=\alpha _d\equiv \alpha$,
$\beta _s=\beta _d\equiv \beta$, $\Delta \mu _s=
\Delta \mu _d\equiv \Delta \mu$, and $T_{cs}=T_{cd}\equiv T_c$.
This situation is realized, for example, in a $d$-wave orthorhombic sample
(like $YBCO$) with twin boundaries which are represented by tetragonal
regions of variable width, with a reduced chemical potential~\cite{14}.
In this particular case, Eqs.(29) and (30) yield for $T<T_c$
\begin{equation}
|\psi _0|^2=\frac{\alpha (T_c-T)-\Delta \mu (\theta )}
{\beta +\Gamma (\theta )},
\end{equation}
where $\Delta \mu (\theta )\equiv \Delta \mu -\Delta (\theta )$.
After substituting the thus found $|\psi _0|$ into Eq.(25) we obtain for the
generalized equilibrium free energy density
\begin{equation}
\Delta \Omega (T,\theta )\equiv \Delta {\cal G} [\psi _0]=-
\frac{[\alpha (T_c-T)-\Delta \mu (\theta) ]^2}{\beta +
\Gamma (\theta )}
\end{equation}
which in turn results in the following expression for the mean-field value
of the thermopower
difference in a $C$-shaped sample with $SND$ junction (see Fig.1)
\begin{equation}
\Delta S_{av}(T,\theta )=S_{p,av}(\theta )-B_{av}(\theta )(T_c-T),
\end{equation}
where
\begin{equation}
S_{p,av}(\theta )=\frac{\beta}{qT_c}\left [\frac{\Delta \mu (\theta )}
{\beta +\Gamma (\theta )}\right ],
\end{equation}
and
\begin{equation}
B_{av}(\theta )=\frac{\alpha}{qT_c}\left [\frac{\beta}{\beta+
\Gamma (\theta )}\right ].
\end{equation}

\subsubsection{Mean-field Gaussian fluctuations of the differential
thermopower: $\Delta S_{fl}(T,\theta )$}

Following the lines of Section II, we can present the fluctuation
contribution to the differential TEP above $(+)$ and
below $(-)$ $T_c$ as
\begin{eqnarray}
\Delta S_{fl}^{\pm}(T,\theta )=A(\theta )[<(\delta \psi _s)^2>_{\pm}
+<(\delta \psi _d)^2>_{\pm}&&\\ \nonumber
+2<\delta \psi _s\delta \psi _d>_{\pm}],&&
\end{eqnarray}
where, e.g.,
\begin{equation}
<(\delta \psi _s)^2>=\frac{1}{Z}\int d|\psi _s|\int d|\psi _d|
(\delta \psi _s)^2e^{-\Sigma [\psi _s,\psi _d]}.
\end{equation}
Here  $Z=\int d|\psi _s|\int d|\psi _d|e^{-\Sigma [\psi _s,\psi _d]}$
is the corresponding partition function with
$\Sigma [\psi _s,\psi _d]\equiv \{\Delta {\cal G} [\psi _s,\psi _d]-
\Delta {\cal G} [\psi _{s0},\psi _{d0}]\}/k_BT$. $A(\theta )$ is a coefficient
to be fixed later.
Expanding the free energy density functional $\Delta {\cal G} [\psi _s,\psi _d]$
\begin{eqnarray}
\Delta {\cal G} [\psi _s,\psi _d]&\approx &\Delta {\cal G} [\psi _{s0},\psi _{d0}] \\ \nonumber
&+&\left[ \frac{\partial ^2\Delta {\cal G}}{\partial \psi _s^2}\right ]_
{|\psi _s|=|\psi _{s0}|}\!(\delta \psi _s)^2 \\ \nonumber
&+&\left[ \frac{\partial ^2\Delta {\cal G}}{\partial \psi _d^2}\right ]_
{|\psi _d|=|\psi _{d0}|}\!(\delta \psi _d)^2 \\ \nonumber
&+&2\left[ \frac{\partial ^2\Delta {\cal G}}
{\partial \psi _s \partial \psi _d}\right ]_{|\psi _n|=|\psi _{n0}|}
\!(\delta \psi _s \delta \psi _d),
\end{eqnarray}
around the mean values of the order parameters $\psi _{n0}$, defined
as stable solutions of equations $\partial {\cal G}/\partial |\psi _n|=0$ we can
explicitly calculate the Gaussian integrals to obtain
\begin{eqnarray}
<(\delta \psi _s)^2>_{-}&=&<(\delta \psi _d)^2>_{-}\\ \nonumber
&=&\frac{k_BT_c\beta}{4(\beta -\Gamma )[\alpha (T_c-T)-\Delta \mu (\theta )]},
\end{eqnarray}
\begin{equation}
<\delta \psi _s\delta \psi _d>_{-}=-
\frac{k_BT_c\Gamma}{4(\beta -\Gamma )[\alpha (T_c-T)-\Delta \mu (\theta )]}
\end{equation}
and
\begin{eqnarray}
<(\delta \psi _s)^2>_{+}&=&<(\delta \psi _d)^2>_{+}\\ \nonumber
&=&\frac{k_BT_c}{2[\alpha (T-T_c)+\Delta \mu (\theta )]},
\end{eqnarray}
\begin{equation}
<\delta \psi _s\delta \psi _d>_{+}=0,
\end{equation}
for the order parameters fluctuations below and above $T_c$, respectively.
In principle, the above expressions completely determine the fluctuation
contribution to the seeking TEP of $SND$ contact in the presence of strong
$N$-metal induced pair-breaking  effects. However, to compare it with the
earlier calculated
mean-field values, let us assume that
$[\Delta \mu (\theta )](E_F/\Delta _0)\gg k_B|T_c-T|$. Then, with a
good
accuracy we can approximate Eqs.(40)-(42) as follows
\begin{equation}
\Delta S_{fl}^{\pm}(T,\theta )\simeq S_{p,fl}^{\pm}(\theta )\pm
B_{fl}^{\pm}(\theta )(T_c-T),
\end{equation}
where
\begin{equation}
S_{p,fl}^{-}(\theta )=-\frac{Ak_BT_c}{2\Delta \mu (\theta )},
\end{equation}
\begin{equation}
B_{fl}^{-}(\theta )=\frac{Ak_BT_c\alpha}{2[\Delta \mu (\theta )]^2},
\end{equation}
and
\begin{equation}
S_{p,fl}^{+}(\theta )=-2S_{p,fl}^{-}(\theta ),
\qquad B_{fl}^{+}(\theta )=2B_{fl}^{-}(\theta ).
\end{equation}
Again, requiring that $S_p^{-}(\theta )=S_p^{+}(\theta )\equiv S_p(\theta )$,
where $S_{p}^{-}=S_{p,av}+S_{p,fl}^{-}$ and $S_{p}^{+}=S_{p,fl}^{+}$, the
above equations give
\begin{equation}
A(\theta )=\frac{2\beta}{3qk_BT_c^2}\frac{[\Delta \mu (\theta )]^2}
{[\beta +\Gamma (\theta )]}
\end{equation}
for the above-introduced parameter (see Eq.(37)).
This in turn leads
to the following expressions for the fluctuation and total contributions to
peaks and slopes through their average counterparts (Cf. Section II):
$S_{p,fl}^{+}=S_p=(2/3)S_{p,av}$, $S_{p,fl}^{-}=-(1/3)S_{p,av}$,
$B_{fl}^{-}=(1/3)B_{av}$, $B_{fl}^{+}=
(2/3)B_{av}$, $B^{-}=B_{av}+B_{fl}^{-}=(4/3)B_{av}$, and $B^{+}=B_{fl}^{+}=
(2/3)B_{av}$. Thus, the ratio $B^{-}(\theta )/B^{+}(\theta )=2$ remains
universal showing no dependence on the relative phase difference $\theta$.
As expected, completely neglecting the interference terms (when both
$\Gamma (\theta ) \to 0$ and $\Delta (\theta ) \to 0$) we recover all the
results of Section II for two independent order parameters. Finally,
the differential TEP of $SND$ junction consisted of two superconductors with
similar critical parameters but markedly different pairing symmetries (like
in a $d$-wave orthorhombic $YBCO$ with $s$-wave tetragonal twin boundaries)
reads
\begin{equation}
\Delta S(T,\theta )=S_{p}(\theta )\pm B^{\pm}(\theta )(T_c-T),
\end{equation}
with
\begin{equation}
S_{p}(\theta )=\frac{2\beta \delta _2}{3qT_c\gamma _2}
\left (\frac{\tilde \delta +\cos \theta}{\tilde \gamma +\cos 2\theta}\right ),
\end{equation}
and
\begin{equation}
B^{-}(\theta )=2B^{+}(\theta )=\frac{4\alpha}{3qT_c\gamma _2}
\left (\frac{\beta}{\tilde \gamma +\cos 2\theta}\right ).
\end{equation}
Here $\tilde \gamma \equiv (\beta +\gamma _1)/\gamma _2$ and
$\tilde \delta \equiv (\Delta \mu +\delta _1)/\delta _2$.
\begin{figure}[htb]
\epsfxsize=8cm
\centerline{\epsffile{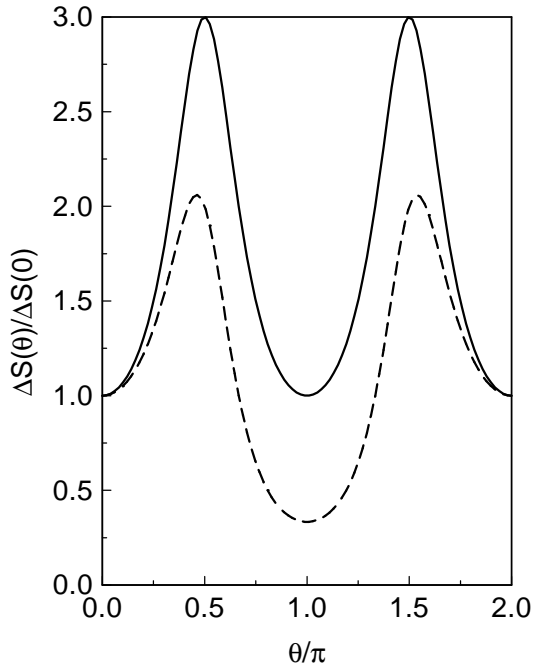} }
\caption{Predicted phase-dependent thermopower response of $SND$
configuration in a $C$-shaped sample (see Fig.1). Solid and dashed lines
depict, respectively, the relative phase $\theta$ dependence of the
normalized slope
$B^{-}(\theta )/B^{-}(0)$ and peak value $S_p(\theta )/S_p(0)$ of the induced
thermopower difference, according to Eqs.(50) and (51).}
\end{figure}
Fig.3 shows the predicted $\theta$-dependent behavior of the normalized
slope $B^{-}(\theta )/B^{-}(0)$ (solid line) and the peak $S_p(\theta )/S_p(0)$
(dashed line) of the $SND$-induced thermopower
difference $\Delta S(T,\theta )$ just below $T_c$, for $\tilde \gamma =
\tilde \delta =2$. As is seen, both the slope
and the peak exhibit a maximum for the $s+id$ state (at $\theta =\pi /2$)
and a minimum for the $s-d$ state (at $\theta =\pi$). Such sharp
dependencies suggest quite
an optimistic possibility to observe the above-predicted behavior of the
induced thermopower, using the sample geometry and experimental technique
described in Section II. Besides, when the pair-breaking  effects (due to
the normal-metal insert in $SND$ junction) are negligible (so that
$\Delta \mu =0$), Eqs.(49)-(51) will describe the differential TEP at the
$SD$ interface where the pair-breaking  interference effects (governed by
the $\Delta (\theta )$ term) will dominate its peak behavior. This situation
would allow one to get a more detailed information about the mixed pairing
states and the introduced phenomenological parameters $\gamma _{1,2}$ and
$\delta _{1,2}$.

\subsection{Low-$T_c$ $s$-wave superconductor and high-$T_c$ $d$-wave
superconductor}

Let us turn now to another limiting case and consider an $SD$ interface
formed by two different superconductors (with
$|\psi _{s0}|\neq |\psi _{d0}|$, $\alpha _s\neq \alpha _d$,
$\beta _s\neq \beta _d$, and $T_{cs}\neq T_{cd}$) in the absence of a
normal-metal layer (which is responsible for pair-breaking  effects).
We shall also assume that the charge-related interference effects (governed
by the $\Delta (\theta )$ term) are rather small and can be safely neglected.
Thus, in this Section we consider the situation when $\Delta \mu _n=0$
and $\Delta(\theta )=0$.
Such a situation can be realized
for an $s$-wave low-$T_c$ superconductor (like $Pb$) and a $d$-wave
high-$T_c$ superconductor (like orthorhombic $YBCO$)~\cite{1,9}.
In fact, the solution for this particular case is well-known. It has been
discussed by Sigrist et al~\cite{11} in a somewhat different context.
In principle, we can obtain both an average and fluctuation contributions to
the resulting TEP for this case, following the recipes of the previous
Section. And in particular, it can be shown that the fluctuation contribution
is still governed by expressions similar to the ones given by Eqs.(40)-(43)
with an evident change in parameters, $\alpha \to \alpha _n$ and
$\beta \to \beta _n$ for $s$- and $d$-wave superconductors. Since,
however, the correponding expressions are rather cumbersome, in what
follows we restrict our analysis with the average values of the induced TEP
only.

Assuming $T_{cs}<T_{cd}$, two temperature regions should be distinguished.

\paragraph{$T<T^{*}(\theta )$.} In this region, the corresponding expressions
for the equilibrium order parameters read (see Eqs.(29) and (30))
\begin{equation}
|\psi _{s0}|^2=\frac{\beta _da_s(T)-\Gamma (\theta )a_d(T)}
{\Gamma ^2(\theta )-\beta _s\beta _d},
\end{equation}
and
\begin{equation}
|\psi _{d0}|^2=\frac{\beta _sa_d(T)-\Gamma (\theta )a_s(T)}
{\Gamma ^2(\theta )-\beta _s\beta _d},
\end{equation}
where the transition point $T^{*}(\theta )$, defined by the equation
$\psi _{s0}(T^{*})=0$, is strongly $\theta$-dependent and deviates from
an $s$-wave critical temperature $T_{cs}$ as follows
\begin{equation}
T^{*}(\theta )=T_{cs}-\frac{\alpha _d \Gamma (\theta ) \Delta T_c}
{\alpha _s \beta _d-\alpha _d \Gamma (\theta )},
\end{equation}
where $\Delta T_c\equiv T_{cd}-T_{cs}$.

After substituting the solution given by Eqs.(52) and (53) into Eq.(25) we
obtain for the average thermopower difference
\begin{equation}
\Delta S_{av}^{I}(T,\theta ;\Delta T_c)=S_{p,av}^{I}(\theta ;\Delta T_c)-
B_{av}^{I}(\theta )[T^{*}(\theta )-T],
\end{equation}
where
\begin{equation}
S_{p,av}^{I}(\theta ;\Delta T_c)=\frac{\alpha _s}{2qN}\left
[\frac{\alpha _d^2\Delta T_c}{\alpha _s \beta _d-
\alpha _d \Gamma (\theta )}\right ],
\end{equation}
and
\begin{equation}
B_{av}^{I}(\theta )=\frac{2\alpha _s\alpha _d\Gamma (\theta )-
\alpha _s^2 \beta _d
-\alpha _d^2\beta _s}{2qN[\Gamma (\theta )^2-\beta _s \beta _d]}.
\end{equation}
Here $N=n_sn_d/(n_s+n_d)$ is a generalized carrier number density.

\paragraph{$T^{*}(\theta )\le T<T_{cd}$.} In this region we obtain from
Eqs.(29) and (30)
\begin{equation}
|\psi _{s0}|=0, \qquad |\psi _{d0}|^2=\frac{\alpha _d(T_{cd}-T)}{\beta _d},
\end{equation}
for the equilibrium order parameters. And the resulting mean-field
thermopower difference in this region is
\begin{equation}
\Delta S_{av}^{II}(T,\theta )=S_{p,av}^{II}(\theta ,\Delta T_c)+
B_{av}^{II}[T-T^{*}(\theta )],
\end{equation}
where
\begin{equation}
S_{p,av}^{II}(\theta ;\Delta T_c)=2S_{p,av}^{I}(\theta ;\Delta T_c), \qquad
B_{av}^{II}=\frac{\alpha _d^2}{qN\beta _d}.
\end{equation}
Figure 4 depicts the ratio $T^{*}(\theta )/T_{cs}$ as a function of
$T_{cd}/T_{cs}$ for different $\theta$.
As we can see, for the chosen set of parameters ($\gamma _1=\gamma _2=
\beta _s$ and $n_d=n_s$), in the mixed $s+id$ pairing state (with
$\theta =\pi /2$, dashed line) $T^{*}(\pi /2)=T_{cs}$ for all $T_{cd}/T_{cs}$.
As it follows from Eqs.(56) and (57), this state is described by the following
dependencies of the TEP peak and slope (below $T^{*}$)
\begin{equation}
S_{p,av}^{I}(\theta =\frac{\pi}{2};\Delta T_c)=\frac{2\Delta _{d0}k_B}{qE_F^d}
\left (1-\frac{T_{cs}}{T_{cd}}\right ),
\end{equation}
and
\begin{equation}
B_{av}^{I}(\theta =\frac{\pi}{2})=\frac{3.52k_B^2}{qE_F^s}\left (1+\frac{E_F^s}{E_F^d}\right ).
\end{equation}
As it is evident from the above equations, in this regime the peak's
amplitude $S_p$ is entirely dominated
by the critical temperatures difference $T_{cd}-T_{cs}$ of the two
superconductors while the slope $B$ is governed by the corresponding
Fermi energies.
It would be interesting to test the predicted behavior of the induced
thermopower at such $SD$ interface using a low-$T_c$ $s$-wave and a
high-$T_c$ $d$-wave superconductors (like, e.g., $Pb$ and $YBCO$).
\begin{figure}[htb]
\epsfxsize=8cm
\centerline{\epsffile{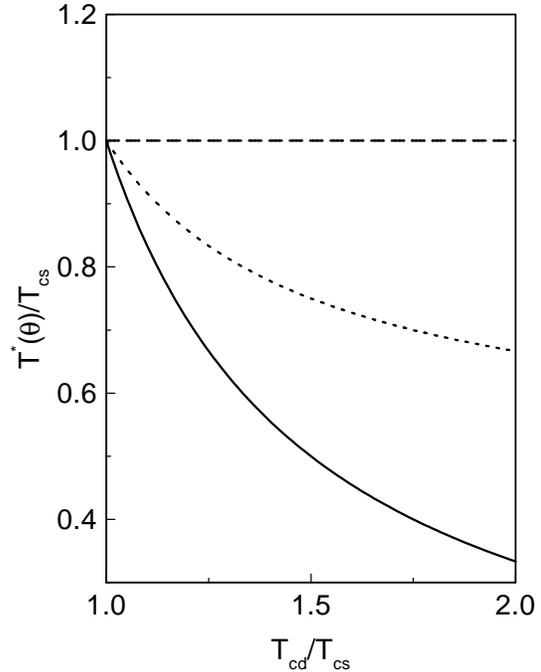} }
\caption{The ratio $T^{*}(\theta )/T_{cs}$ as a function of $T_{cd}/T_{cs}$
for different $\theta $ calculated according to Eq.(54).
Solid, dashed, and dotted lines correspond to $\theta =\pi $ ($s-d$ state),
$\theta =\pi /2$ ($s+id$ state), and $\theta =\pi /4$, respectively.}
\end{figure}

\section{Conclusion}

In summary, to probe into the pairing state of high-$T_c$
superconductors, we calculated the differential thermopower $\Delta S$ of
$SND$ junction in the presence of strong
pair-breaking  effects (due to the normal-metal layer $N$) and charge-related interference
effects (due to the chemical imbalance at $SD$ interface) using the
generalized Ginzburg-Landau theory for an admixture of $s$-wave and $d$-wave
superconductors near $T_c$.
The calculated thermopower was found to strongly depend on the relative
phase $\theta =\phi _s-\phi _d$ between the two superconductors
(exhibiting a pronounced maximum near the mixed $s+id$ state with
$\theta =\pi /2$) and their critical temperatures. The experimental
conditions under which the predicted
behavior of the induced thermopower could be observed were discussed.

\acknowledgments

We thank J. Annett, J. Clayhold and T.M. Rice for their interest in this
work and very useful discussions.
S.S. was financially supported by FNRS (Brussels, Belgium).
M.A. was financially supported by the Minister of Education under contract
No. ARC (94-99/174) of ULg.

\end{document}